\documentclass[a4paper,
               ]{jacow}
\ifboolexpr{bool{xetex} or bool{luatex}} 
 {}                                      
 {\usepackage[utf8]{inputenc}}           

\usepackage[USenglish]{babel}

\begin{document}

\title{Optimization of piezo operation for superconducting TESLA cavities at \NoCaseChange{EuXFEL}}

\author{M. Grecki\thanks{mariusz.grecki@desy.de}, 
		J. Branlard,
		M. Diomede,
		M. Herrmann, 
		C. Schmidt \\
		Deutsches Elektronen-Synchrotron DESY, Hamburg, Germany
		}
	
\maketitle

%

\begin{abstract}

Superconducting cavities with high Q-factor require precise tuning to match the RF frequency, ensuring stable electromagnetic fields and minimizing RF power consumption. At the European XFEL (EuXFEL) accelerator, TESLA cavities are tuned using slow tuners (step motors) for coarse adjustments and fast tuners (piezoelectric actuators) for fine-tuning and compensating disturbances such as Lorentz Force Detuning (LFD) and microphonics.

Piezo actuators, critical to this system, require high-voltage (up to 100\,V) and high-current (up to 1\,A) driving signals for effective LFD compensation. However, they are vulnerable to overvoltage, overcurrent, and overheating, and their protection is crucial since replacing damaged piezos in fully assembled modules is infeasible. Additionally, piezo-induced vibrations can affect the machine's stability.

Optimizing piezo excitation—by reducing voltage, current, and current slope while ensuring effective LFD compensation—improves equally efficiency, reliability and machine stability. This paper explores the optimization of piezo excitation at EuXFEL, detailing methods and results applicable to other facilities with superconducting cavities as well.

\end{abstract}
%

\section{Introduction}
\begin{figure}
\includegraphics[width=\columnwidth]{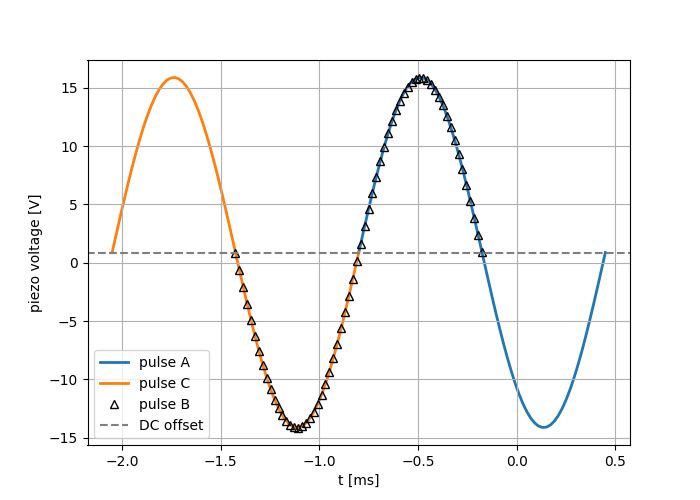}
\caption{Three sample single-period piezo pulses (A, B, and C) from the parameter search space.}
\label{fig:pulse_setup}
\end{figure}

The cavities at the EuXFEL~\cite{xfel} are tuned using a combination of stepper motor tuners and piezoelectric actuators. Stepper motors provide coarse tuning, while piezo actuators enable fine tuning and compensation of Lorentz force detuning (LFD). Various strategies for piezo actuation have been proposed and implemented~\cite{alg_fnal, piezo_xfel}. At EuXFEL, a simple and robust approach based on the generation of sinusoidal voltage pulses is employed (Fig.~\ref{fig:pulse_setup}). These pulses are generated with carefully adjusted amplitude and timing to counteract cavity deformation induced by RF power. The primary advantage of this method lies in its simplicity and operational reliability.

\begin{figure}
\centering
\includegraphics[width=\columnwidth]{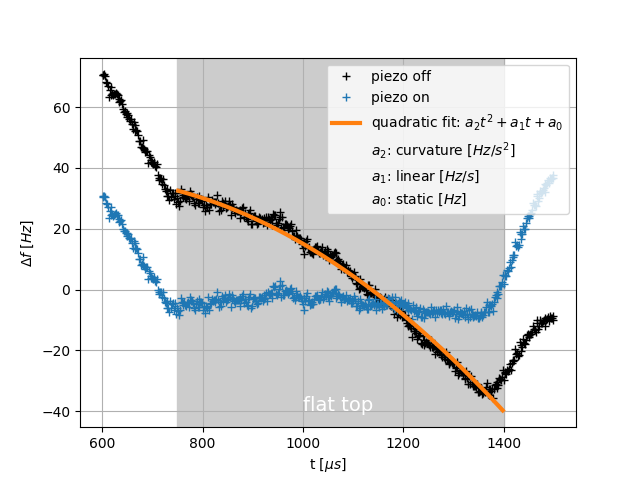}
\caption{Detuning profile over the RF pulse and definition of detuning parameters \textit{static}, \textit{linear}, and \textit{curvature} with piezo automation off (black) and on (blue).}
\label{fig:detuning}
\end{figure}

The key parameters of the piezo pulse—amplitude, frequency, timing relative to the RF pulse, and DC offset—can be adjusted. The excitation frequency is fixed at approximately 300\,Hz, corresponding to the mechanical resonance frequency of the cavity. Cavity detuning during the RF pulse—specifically during the flattop phase when the field is stabilized—is characterized by three components: static, linear, and quadratic (curvature) detuning (Fig.~\ref{fig:detuning}). A PI control scheme is used to iteratively adjust the piezo pulse parameters in order to minimize these detuning components: the pulse amplitude compensates for linear detuning, the DC offset corrects static detuning, and the pulse timing addresses curvature-related detuning effects (Fig.~\ref{fig:detuning}).

An automated tuning procedure, based on this algorithm, has been implemented to achieve near-zero detuning during the RF flattop. The effectiveness of this method is demonstrated in Fig.~\ref{fig:pulse_setup}, which shows the detuning waveform with active piezo compensation.

The operational lifetime of piezoelectric actuators is influenced by both environmental conditions and excitation parameters. Specifically, temperature, humidity, and applied voltage have been identified as critical factors~\cite{piezo_reliability1, piezo_reliability2}. In the cryogenic environment of the EuXFEL, temperature and humidity can be assumed constant, leaving excitation voltage as the primary factor affecting long-term reliability.

As the lifetime of piezo actuators decreases exponentially with increasing operating voltage, minimizing the excitation voltage is essential for maximizing durability. Furthermore, operational experience indicates that reducing the current slope at the onset of the piezo pulse helps to prevent localized inrush currents and reduces the risk of thermal runaway caused by positive electrothermal feedback mechanisms.

The following sections present an optimization strategy for piezo pulse parameters, with the dual objective of minimizing excitation voltage and reducing the initial current slope of the piezo pulse.

%

\section{optimization of \\ Piezo pulse parameters }

To identify the conditions under which the excitation voltage required to compensate Lorentz force detuning (LFD) is minimized, we performed a systematic characterization of various piezo drive waveforms across all 32 cavities powered by a single RF station (A23.L3 at EuXFEL). Under constant RF excitation conditions (650\,MV in vector sum, with a standard EuXFEL RF pulse shape), each cavity was tuned using different piezo pulse configurations.

Several actuation strategies were explored, varying the piezo pulse frequency, timing relative to the RF pulse, and phase (Fig.~\ref{fig:pulse_setup}, examples A, B, and C). Additionally, double-period sinusoidal pulses were tested.

An automated optimization routine was employed to scan the frequency dependence of piezo pulse parameters required for LFD compensation. For each pulse frequency, the cavity was tuned by iteratively adjusting the pulse amplitude, timing, and DC offset voltage to minimize the detuning components during the RF flattop (Fig.~\ref{fig:detuning}). The objective was to reduce the static, linear, and quadratic detuning components to near zero; the optimization was considered complete once the absolute values of all three detuning parameters fell below 10\,Hz. At this point, the required piezo pulse amplitude was recorded for further analysis. The DC offset voltage was not considered critical, as it can be compensated using the stepper motor tuners.

\section{Piezo Signal Smoothing}

\begin{figure}
\includegraphics[width=\columnwidth]{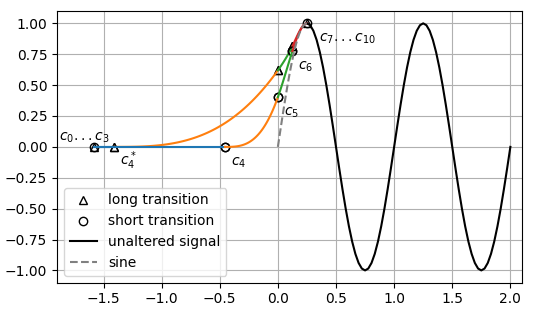}
\caption{Comparison of piezo voltage signals. $\textit{dashed}$: sine, $\Delta$: control points of cubic spline with long transition time, o: control points of cubic spline with short transition time. The control points are denoted $c_0 \hdots c_{10}$. Overlapping control points are at the start and end of the respective spline. For better readability, only control points of the short transition spline are listed. Control point $c_4$ is the single free parameter used for optimization where $c_4$ corresponds to the short transition spline, and $c_4^*$ corresponds to the long transition time, respectively.}
\label{fig:sine_vs_spline}
\end{figure}

Smoothing is done by cutting the initial part of the sine signal, and substituting it with a cubic spline~\cite{lyche2008}. Formally put,

\begin{equation}
  u(t) =
  \begin{cases}
    p(t) & \text{if} \qquad -T_0 \le t \le T/4 \ , \\
    \sin(\omega t) & \text{if} \qquad T/4 < t \le 2T \ , \\
    0 & \text{otherwise} \ ,
    \end{cases}
  \end{equation}
where $p$ denotes the spline function, $T_0 + T/4$ is the spline transition time, and $T$ is a sine period. Figure~\ref{fig:sine_vs_spline} illustrates the difference between a default sine pulse, and a compound one.

The requirements for the spline part of the compound piezo voltage signal are as follows:
\begin{enumerate}
\item Smooth up to the $2^{\text{nd}}$ order at the interconnection point of sine and spline.
\item Smooth up to the $2^{\text{nd}}$ order at the beginning of the pulse.
\item One free parameter for shaping the initial guess.
\end{enumerate}

These requirements form $n_c=7$ constraints that require the use of $n_c - d = 4$ interconnected splines, where $d=3$ is the order of the polynomial.

The knots $t_i$, i.e. the points in time where we require that $p(t_i) = c_i$, are a design choice for the initial guess of the spline curve, and set by trial and error (see Fig.~\ref{fig:sine_vs_spline}).

Knot $t_4$ was chosen as an optimization parameter, or decision variable, to minimize the cost function

\begin{equation}\label{eq:cost}
J = w \cdot T_0^2(t_4) + (1 - w) \cdot \left(\max\left(p_t(t_4)\right)\right)^2 \ ,
\end{equation}
where $T_0$ is the transition time, or duration of the spline, $p_t$ the time derivative of the spline within the range $[t_6, t_7]$, and $w \in [0, 1]$ a weighting coefficient, where $w = 1$ corresponds to a signal closest to the sine, and $w = 0$ to a smooth spline with maximum transition time. The optimized piezo voltage signal was considered a trade-off between transition time and maximum inrush value, and the weighting coefficient chosen to be $w=0.5$.

\section{results}

\begin{figure}
	\includegraphics[width=\columnwidth]{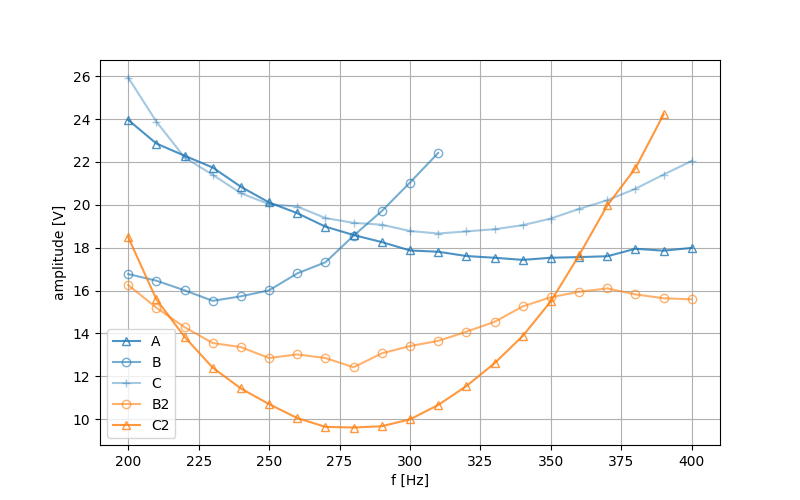}
	\caption{The required piezo pulse amplitude to compensate Lorentz force detuning (LFD) evaluated for three different single-period sinusoidal pulses (labeled A, B, and C, as defined in Fig.~\ref{fig:pulse_setup}) and two corresponding double-period pulses (B2 and C2).}
	\label{fig:frequency}
\end{figure}

\begin{figure}
\includegraphics[width=\columnwidth]{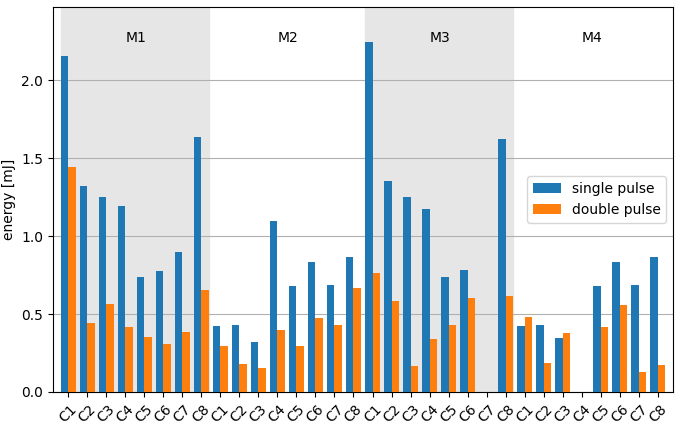}
\caption{Reduction of the total required piezo pulse energy achieved by applying a double-period sinusoidal waveform with optimized frequency and timing (EuXFEL A23.L3).}
\label{fig:energy}
\end{figure}

The results of the piezo pulse parameters optimization are summarized in  Figs.~\ref{fig:frequency} and~\ref{fig:energy}. The key conclusions are as follows:

    Different pulse configurations can significantly affect the required piezo pulse amplitude, with variations reaching several tens of percent.

    While the optimal excitation frequency may vary depending on the pulse shape and timing, it shows minimal dependence on the specific characteristics of individual piezos. Therefore, to simplify implementation and ensure standardization, a representative frequency of 275\,Hz has been selected as the default excitation frequency.

    Among the tested strategies, double-period sinusoidal pulses were found to be the most effective. They enable a reduction in the required piezo amplitude by approximately a factor of two. Since the injected energy scales with the square of the amplitude, this translates to a fourfold reduction in energy input. Taking into account the doubled duration of the pulse, the net energy reduction is approximately a factor of two.

For the piezo signal smoothing, substituting the default sine signal with the smooth signal caused a reduction of the inrush by about $80\%$, as listed in Table~\ref{tab:results}. Settling time and energy consumption are almost the same.

\begin{table}
  \caption{Comparison of Figures of Merits of Piezo Signal Smoothing for 32 Cavities of Station A23 at EuXFEL}
  \setlength\tabcolsep{3pt}
  \begin{tabular}{lll} \toprule
    & \textbf{w/o smooth start} & \textbf{w/ smooth start} \\ \midrule
    settling time [s] & 119.3 ($\pm$ 2.6) & 119.9 ($\pm 0.4$) \\
    energy [mW] & 0.71 ($\pm$ 0.45) & 0.71 ($\pm$ 0.46) \\
    max. inrush [mA/s] & 200 ($\pm$ 63) & 33 ($\pm$ 8) \\ \bottomrule
  \end{tabular}
  \label{tab:results}
  \end{table}

%
%

\section{Conclusion}
Shifting from a conventional single sine voltage to a double sine waveform, combined with spline-based smoothing of the signal’s leading edge, has significantly improved piezo control efficiency. This transition cuts energy consumption by half and reduces inrush current by \qty{80}{\%}, effectively lowering electrical stress and enhancing system stability and sustainability. Importantly, the change requires no hardware modification, no modification of the existing PI control parameters, and the closed-loop settling time remains unaffected.
This approach has already been successfully implemented at EuXFEL, where it demonstrated no negative impact on machine performance while halving the energy required for piezo excitation. Additional smoothing of the inrush current will be introduced in the next version of the piezo control server.

%
%
%

\section{acknowledgements}

This work was partially funded in the context of the R\&D program of the European XFEL. This project has also received funding from the European Unions Horizon Europe research and innovation program under grant agreement N.~101131435

\end{document}